\documentclass[prb,aps,twocolumn,amsmath,amssymb,floatfix,superscriptaddress]{revtex4-1}
\usepackage[dvips]{graphics}
\usepackage{color}
\definecolor{dred}{rgb}{0,0,0.6}
\usepackage{graphicx}  
\usepackage{dcolumn}   
\usepackage{bm}        
\usepackage{amssymb}   
\usepackage{amsmath}
\usepackage{braket}
\usepackage[mathscr]{euscript}
\usepackage{color}
\usepackage{hyperref}
\begin{document}

\title{Photo-induced directional transport in extended SSH chains}

\author{Usham Harish Kumar Singha}
\email{harishsingha07@gmail.com}
\affiliation{Department of Physics, School of Applied Sciences, University of Science and Technology Meghalaya, Ri-Bhoi-793 101, India}

\author{Kallol Mondal }
\email{kallolsankarmondal@gmail.com}
\affiliation{AGH University of Krakow, Faculty of Physics and Applied Computer Science, Aleja Mickiewicza 30, 30-059 Krakow, Poland}

\author{Sudin Ganguly}
\email{sudinganguly@gmail.com}
\affiliation{Department of Physics, School of Applied Sciences, University of Science and Technology Meghalaya, Ri-Bhoi-793 101, India}

\author{Santanu K. Maiti}
\email{santanu.maiti@isical.ac.in}
\affiliation{Physics and Applied Mathematics Unit, Indian Statistical Institute, 203 Barrackpore Trunk Road, Kolkata-700108, India}

\date{\today}
\begin{abstract}
We investigate the current-voltage characteristics of an extended Su-Schrieffer-Heeger (SSH) chain under irradiation by arbitrarily polarized light, demonstrating its potential as a light-controlled rectifier. Irradiation of light induces anisotropy in the system, enabling directional current flow and active control of rectification behavior. Our analysis demonstrates that, under optimized light parameters, the rectification efficiency can exceed 90\%. Moreover, the direction of rectification--whether positive or negative--can be precisely controlled by varying the polarization of the light, highlighting the potential for external optical control of electronic behavior. The effect of light irradiation is incorporated using the Floquet-Bloch ansatz combined with the minimal coupling scheme, while charge transport is computed through the nonequilibrium Green's function formalism within the Landauer-B\"{u}ttiker framework. 

\end{abstract}
\maketitle

\section{\label{sec1}Introduction}
Current rectification is a fundamental operation in electronic circuits, characterized by asymmetric current flow under opposite bias polarities (i.e., $I(V) \neq I(-V)$). Traditionally, this function has been carried out using bulk semiconducting materials. However, recent advancements in nanoscale measurement techniques have significantly transformed this landscape. The reliance on conventional bulk semiconductors is increasingly being replaced by nanoscale systems, which offer substantially enhanced rectification performance due to their improved structural precision and electronic properties~\cite{AVIRAM1974277}.  Presently, the realization of rectifying behavior at the nanoscale has predominantly relied on molecular assemblies or multi-molecular systems~\cite{Ashwell20047102,Diez-Perez2009,Nijhuis2010,Yee2011,Batra2013,Yoon2014}, where asymmetric charge transport is achieved through tailored electronic structures or donor-acceptor configurations. However, the fabrication of efficient single-molecule rectifiers remains a formidable challenge due to limitations in synthetic control, electrode-molecule coupling, and the intrinsic complexity of charge transport mechanisms at the molecular level~\cite{Janes2009}. Consequently, the development of rectifiers based on structurally simple, geometrically asymmetric systems that exhibit high rectification ratios~\cite{Majhi2022,D2CP03823D} has emerged as a compelling direction, offering potential advantages in scalability, reproducibility, and integration into nanoscale electronic architectures.

To achieve rectification, characterized by an asymmetric current--voltage response, the energy level alignment of the funtional element must differ under forward and reverse bias conditions~\cite{Xiang2016}. This asymmetry can be introduced primarily through two strategies~\cite{AVIRAM1974277,10.1063/1.471396,MUJICA2002147,10.1063/1.120195,PhysRevB.64.085405,PhysRevB.66.165436,PhysRevB.70.245317,PhysRevB.73.245431,PATRA2019408}. In the first approach, a structurally symmetric conductor is asymmetrically coupled to the electrodes, resulting in bias-dependent transmission characteristics~\cite{PhysRevB.70.245317,PhysRevB.73.245431,Batra2013,KangShinLee2017}. In the second approach, a structurally asymmetric conductor is symmetrically coupled to the contacts~\cite{10.1063/1.120195,MUJICA2002147,10.1063/1.471396,PhysRevB.66.165436}. The latter method is more widely adopted, as it generally leads to enhanced rectification performance due to the intrinsic asymmetry in the electronic structure of the conductor.

Alongside structural asymmetry, irradiation provides a powerful route to tune electronic structure through Floquet engineering. The foundations of Floquet theory were laid in the study of time-periodic Hamiltonians and driven tunneling~\cite{sambe1973steady,grifoni1998driven}. Light-induced topology emerged with the photovoltaic Hall effect in graphene and the proposal of Floquet topological insulators~\cite{Oka2009,Kitagawa2010,Lindner2011}, followed by theoretical developments of Floquet-Bloch band structures, including Dirac point merging and topological transitions~\cite{PhysRevLett.110.200403,PhysRevB.88.245422}. Experimental realization in topological insulators further confirmed Floquet-Bloch states~\cite{Wang2013}. Later works extended these ideas to driven SSH chains, addressing edge states, long-range hopping, and disorder~\cite{Thakurathi2013,PerezGonzalez2019}. Recent reviews and studies highlight Floquet band engineering, polarization control, and solitonic excitations~\cite{Rudner2020,Seshadri2022,Mandal2024}.

Conventional rectification strategies rely on structural asymmetry or asymmetric coupling. On the other hand, Floquet studies have mainly addressed polarization-dependent effects in topological systems. In this work, we adopt a different approach. We consider a structurally symmetric extended SSH chain under light irradiation and investigate its rectification response. It is well-established that irradiation with arbitrarily polarized light can induce spatial asymmetry~\cite{PhysRevB.88.245422,10.1063/5.0213895} in otherwise symmetric systems, thereby enabling current rectification. Furthermore, the application of light not only breaks spatial symmetry but also provides a means to externally control the rectification effect~\cite{PhysRevLett.110.200403,Rudner2020}. By tuning the applied field parameters,  the rectification properties can be modulated with a high degree of flexibility, offering a versatile platform for optoelectronic control in low-dimensional systems.  A simple SSH model, composed of alternating nearest-neighbor hoppings, retains inversion symmetry (centrosymmetric) and typically does not exhibit rectification when symmetrically coupled to electrodes. In such cases, even under light irradiation, the symmetric band structure limits the generation of directional current. To overcome this, we employ an extended SSH model that incorporates hopping trimerization, in contrast to the dimerized hopping of the conventional SSH model~\cite{perezgonzalez2019}. This trimerized structure breaks inversion symmetry and enables the system to respond non-trivially to external driving fields. As a result, the extended SSH chain offers a more effective platform for realizing and enhancing rectification effects.  In addition to hopping trimerization, the geometric arrangement of the unit cell is crucial for determining the rectification response under irradiation. In a linear chain, the effect of the external field is confined to a single direction, so the light effectively acts as linearly polarized irrespective of its actual polarization. This restricts the scope for polarization-based control. In contrast, a zigzag geometry (see Fig.~\ref{Fig1}) couples to multiple field components, allowing the polarization state of the light to directly modulate the hopping amplitudes. This enhanced tunability makes the zigzag configuration particularly attractive, as rectification can now be controlled not only by field strength and frequency but also by polarization. Together, trimerized hopping and zigzag geometry provide a versatile framework where structural asymmetry and light–matter interaction cooperate to achieve enhanced rectification.

Therefore, we consider an extended SSH chain with trimerized bonds arranged in a zigzag pattern and subjected to arbitrarily polarized light. The system is modeled within the tight-binding framework, with irradiation incorporated through the standard Floquet-Bloch ansatz in the minimal coupling scheme~\cite{PhysRevLett.110.200403,sambe1973steady,grifoni1998driven,PhysRevB.88.245422}. Transport properties are then computed using the well-established Green's function formalism within the Landauer-B\"{u}ttiker approach~\cite{datta1997electronic,datta2005quantum,di2008electrical,nikolic2000quantum}.

The key findings of this work are (i) light irradiation induces spatial asymmetry in the system, resulting in pronounced rectification behavior, (ii) the rectification efficiency is significantly enhanced under optimal irradiation conditions, and
(iii) the rectification characteristics can be tuned by adjusting the parameters of the applied field. These findings highlight the possibility of the system as a flexible platform for optoelectronic control in low-dimensional systems.

The rest of the presentation is organized as follows. In Sec.~\ref{sec2}, we present the model and the necessary theoretical formulation to compute current rectification ratio. All the results are critically analyzed in Sec.~\ref{sec3}. Finally, in Sec.~\ref{con}, we end with concluding remarks. A detailed derivation of the Floquet formalism used to construct the Hamiltonian in the presence of light is provided in Appendix~\ref{appa}. The band structure analysis is presented in Appendix~\ref{appb}, and the frequency-dependent energy spectrum is discussed in Appendix~\ref{appc}.
\section{\label{sec2}Model and theoretical formulation}

\begin{figure}[ht!]
\includegraphics[width=0.48\textwidth]{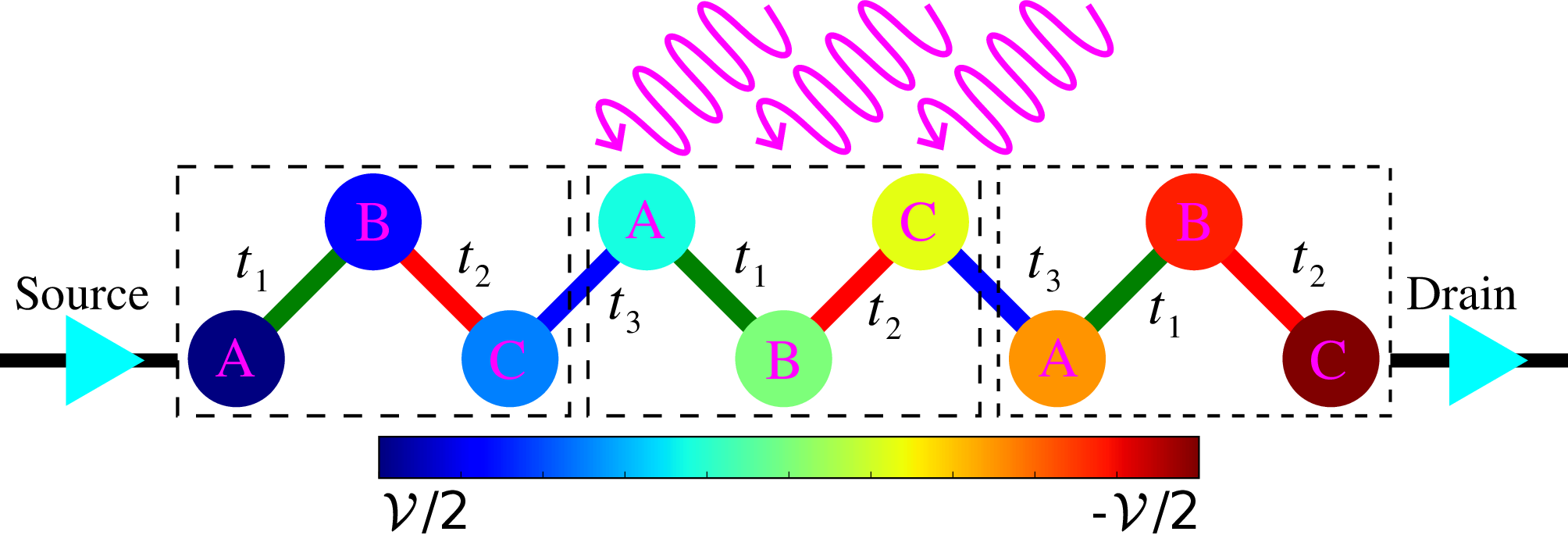}
\caption{(Color online.) Schematic diagram of an extended SSH chain in presence of light irradiation, coupled to two 1D electrodes, source and drain. These electrodes are semi-infinite, metallic, and non-magnetic in nature. Each unit cell of the chain is indicated with a dotted box and composed of three sites. $t_1$ and $t_2$ are the intracell hopping integrals denoted with the green and red bonds, while $t_3$ is the the intercell hopping denoted with blue bonds. The different site colors represent the linear variation of on-site energies under the applied bias ${\mathcal V}$, as indicated by the horizontal colorbar. The irradiation effect is represented with the magenta waves.}
\label{Fig1}
\end{figure}

Our proposed setup is schematically illustrated in Fig.~\ref{Fig1}. Here, an extended SSH chain, is connected to two electrodes: the source and the drain. These electrodes are modeled as one-dimensional, 
metallic, and semi-infinite. The bare system is exposed to arbitrarily polarized light, while the electrodes remain unaffected by any irradiation or interactions. A finite bias voltage ${\mathcal V}$ is applied across the junction, resulting in the modification of the site energies.

\subsection{Model Hamiltonian}
We model the system within the nearest-neighbor tight-binding framework, where the corresponding Hamiltonian is expressed using second quantization.
The  total Hamiltonian $H$ is composed of four parts and can be written as
\begin{equation}
H = H_\text{SSH}  + H_\text{S}  + H_\text{D}  +H_\text{C},
  \label{Eqn0}
\end{equation}
where $H_\text{SSH}$, $H_\text{S}$, $H_\text{D}$  and $H_\text{C}$ are the sub-parts of the Hamiltonian associated with the extended SSH chain, source, drain, and the coupling between the electrodes and the molecular junction, respectively. The Hamiltonian for the central region is given by~\cite{adamatios,ghun}
\begin{align}
 &H_\text{SSH}  = \sum_{n=1}^N \sum_\alpha\epsilon_{n,\alpha} c_{n,\alpha}^\dagger c_{n,\alpha}   \nonumber \\
  & + \sum_{n=1}^{N}  \left[t_1 \left( c_{n,A}^\dagger ~c_{n,B} + \text{h.c.} \right) +  t_2 \left( c_{n,B}^\dagger ~c_{n,C} + \text{h.c.} \right)\right] \nonumber \\
  & +  \sum_{n=1}^{N-1} t_3 \left( c_{n,C}^\dagger ~c_{n+1,A} + \text{h.c.} \right) ,
  \label{Eqn1}
\end{align}

\noindent
where $c_{n,\alpha}/c^\dagger_{n,\alpha}$ annihilates/creates a particle of type $\alpha = A, B, C$ in the unit cell $n = 1, \ldots, N$. $\epsilon_{n,\alpha}$ represents the on-site potential at the $\alpha$th sublattice in the $n$th unit cell.  $t_1$ and $t_2$ are the intracell and $t_3$ is the intercell hopping amplitudes in the absence of light. 

The source and drain electrodes attached to the SSH chain are modeled as one-dimensional, perfect, semi-infinite, and described by the standard nearest-neighbor tight-binding Hamiltonian~\cite{PhysRevB.108.195401,D2CP02523J}. The sub-Hamiltonians $H_\text{S}$, $H_\text{D}$, and  $H_\text{C}$ are
\begin{eqnarray}
H_{\rm S} &=& H_{\rm D} = \epsilon_0\sum\limits_{\alpha} d_\alpha^{\dagger} d_\alpha +
t_0\sum\limits_{\langle \alpha \beta\rangle}\left(d_\alpha^{\dagger} d_\beta + h.c.\right),\\
H_{\rm C}  &=& H_{\rm S,\rm SSH} + H_{\rm D, \rm SSH} \nonumber \\
& = & \tau_S\left(c_{1,A}^{\dagger} d_0 + h.c.\right) + \tau_D\left(c_{N,C}^{\dagger} d_{N+1} + h.c.\right),
\label{lead-ham}
\end{eqnarray} 
where $\epsilon_0$ and $t_0$ represent the on-site energy and hopping amplitude of the source and drain electrodes, respectively. The operators $d_\alpha$ and $d_\alpha^\dagger$ correspond to the annihilation and creation operators at site $\alpha$, respectively in the electrodes. The notation $\langle \cdots \rangle$ indicates that the interaction is restricted to nearest-neighbor sites only. The coupling matrix $H_\text{C}$ consists of two parts, the coupling between the source and the extended SSH chain, denoted by $H_{\rm S,\rm SSH}$, and the coupling between the chain and the drain, denoted by $H_{\rm D,\rm SSH}$. The source and drain are connected to the central region via the coupling parameters $\tau_S$ and $\tau_D$, respectively.   


\subsection{Inclusion of bias voltage}
When a finite bias voltage is applied between the source and drain electrodes, an electrostatic potential develops across the system, modifying the on-site energies of each of the  sites, while those in the electrodes remain unaffected~\cite{10.1063/1.1539863}. This spatial variation in potential can be incorporated into the system Hamiltonian through the voltage-dependent on-site energy term, which is typically decomposed into two components as
\begin{equation}
\epsilon_{n,\alpha} = \epsilon_{n,\alpha}^0 + \epsilon_{n,\alpha}({\mathcal V}),
\end{equation}
where \( \epsilon_{n,\alpha}^0 \) is the voltage-independent contribution arising from the intrinsic properties of the material, which may be either randomly or deterministically disordered. The voltage-dependent component \( \epsilon_{n,\alpha}({\mathcal V}) \) accounts for the influence of the applied bias, including both the bare electric field and screening effects due to electron redistribution. 

In the absence of significant screening, \( \epsilon_{n,\alpha}({\mathcal V}) \) generally follows a linear variation along the transport direction within the molecular region, while remaining constant across the transverse direction. Although various forms of potential profiles have been considered in the literature, the linear profile remains the most widely used due to its simplicity and reasonable physical approximation~\cite{10.1063/1.1539863,Guo2016ls,Saha2019,Majhi2022}. In this work, we also adopt a linear potential profile across the central system. For a linear bias profile ${\mathcal V}$, the on-site potential is given by
\begin{equation}
    \epsilon_{n,\alpha} = {\mathcal V}\left[\frac{1}{2} - \frac{3(n-1) + (\alpha-1)}{3N-1}\right],
\end{equation}
where $n = 1,2,\ldots,N$ and $\alpha = 1,2,3$
denoting the sublattices $A$, $B$, and $C$, respectively.


\subsection{Incorporation of light irradiation}
The effect of light irradiation is included through the hopping terms in the Hamiltonian. When the system is exposed to light, it becomes periodically driven. Using the minimal coupling scheme and the Floquet–Bloch ansatz~\cite{PhysRevLett.110.200403,sambe1973steady,grifoni1998driven,PhysRevB.88.245422}, the effect of light can be incorporated via a vector potential $\mathbf{A}(t)$. In the system Hamiltonian, the vector potential affects the hopping integral through the minimal coupling scheme under the dipole approximation~\cite{PhysRevLett.110.200403} $\mathbf{k}\Rightarrow \mathbf{K}(\tau) = \mathbf{k} + q\mathbf{A}(\tau)/\hbar$, where $q$ is the electronic charge, $\mathbf{A}(\tau)$ is the vector potential of the external light field, and $\hbar$ is the reduced Plank constant. Generally, the vector potential can be expressed as $\mathbf{A}(\tau) = ({\mathcal A}_x \sin(\Omega \tau), {\mathcal A}_y \sin(\Omega \tau + \phi), 0)$, representing an arbitrarily polarized field in the \(x\)–\(y\) plane. Here, \({\mathcal A}_x\) and \({\mathcal A}_y\) are field amplitudes, and \(\phi\) denotes the phase of the polarized field. Different choices of \({\mathcal A}_x\), \({\mathcal A}_y\), and \(\phi\) can achieve various polarized lights, such as circularly, linearly, or elliptically polarized lights. We choose the dimensionless parameters $A_x = e{\mathcal A}_x a/\hbar$ and $A_y = e{\mathcal A}_y a/\hbar$ to charaterize the driving field. Therefore, with $A_x$ and $A_y$ of order unity, the magnitude of the vector potential is approximately $10^{-6}\,$Tesla-meter. This justifies the use of the dipole approximation, since the wavelength employed in the present study is about $300\,\text{nm}$ (mentioned in the Results section), whereas the system dimension is only a few nanometers. Consequently, the spatial variation of the vector potential across the sample is negligible, and the dipole approximation is well justified. In the presence of irradiation, the effective hopping integral takes the form~\cite{PhysRevLett.110.200403,Mondal2021,GANGULY2021302,10.1063/5.0045566},
\begin{equation}
\tilde{t}_{nm}^{p,q} = t_{nm} \cdot \frac{1}{\mathbb {T}} \int_0^\mathbb {T} e^{i\Omega \tau (p-q)} e^{i \mathbf{A}(\tau) \cdot \mathbf{d}_{nm}} d\tau,
\label{effhop}
\end{equation}
where the vector $\mathbf{d}_{nm}$ connects the nearest-neighbor sites of the system. $t_{nm}^{pq}$ is the modified hopping amplitude in the presence of light. The indices in the suffix $n$ and $m$ represent the nearest-neighbor connections between sites $n$ and $m$. The indices $p$ and $q$ refer to the Floquet bands. The driving field is considered uniform with a frequency $\Omega$ and a period $\mathbb {T}$.

Let $\mathbf{d}_{nm} = d_x \hat{\bf x} + d_y \hat{\bf y}$ represent the position vector between sites $n$ and $m$, and with the vector potential $\mathbf{A}(\tau)$ for an arbitrarily polarized light, the time-averaged hopping amplitude from Eq.~\ref{effhop} simplifies to 
\begin{eqnarray}
\tilde{t}_{nm}^{p,q} &=&  \frac{t_{nm}}{\mathbb{T}}\int_0^\mathbb{T} e^{i \Omega \tau(p-q)} e^{iA_xd_x\sin{\Omega\tau}} e^{iA_yd_y\sin{\left(\Omega\tau +\phi\right)}} d\tau\nonumber\\
&=&t_{nm} e^{i (p-q)\Theta } J_{(p-q)}\left(\Gamma\right) 
\label{effhop1}
\end{eqnarray}
where the parameters $\Gamma$ and $\Theta$ are given by
\begin{eqnarray}
\Gamma &=& \sqrt{\left(A_xd_x\right)^2 + \left(A_yd_y\right)^2 + 2A_xA_yd_xd_y\cos{\phi}}\label{gam}\\
\Theta &=& \tan^{-1}{\left(\frac{A_yd_y\sin{\phi}}{A_xd_x + A_yd_y\cos{\phi}}\right)}.\label{thet}
\end{eqnarray}
Here $J_{(p-q)}$ denotes the Bessel function of the first kind of order $(p-q)$. As evident from Eq.~\ref{effhop1}, the effective nearest-neighbor hopping amplitude becomes direction-dependent. This directional sensitivity in the hopping terms allows the external field to induce spatial anisotropy in the chain. The Floquet formalism to get Eq.~\ref{effhop1} is given in Appendix~\ref{appa}.
\subsection{Formulation of rectification performance}
To analyze rectification performance, we define the rectification ratio (RR) as~\cite{Saha2019,Majhi2022}
\begin{equation}
RR = \frac{|I(+{\mathcal V})| - |I(-{\mathcal V})|}{|I(+{\mathcal V})| + |I(-{\mathcal V})|} \times 100\%.
\end{equation}
Here $I(+{\mathcal V})$ and $I(-{\mathcal V})$ are the forward bias (FB) and reverse bias (RB) currents at voltage ${\mathcal V}$. An $RR = 0$ implies no rectification, while an $RR = \pm 100\%$ indicates maximum current rectification. A positive sign indicates that the forward bias current dominates over the reverse bias current, while a negative sign signifies the opposite.

The calculation of current begins with the evaluation of the two-terminal transmission probability, which we compute using the nonequilibrium Green's function (NEGF) technique within the Landauer--B\"{u}ttiker framework. The retarded Green's function is defined as
\begin{equation}
G^r(E) = \left[ E{\mathbb I} - H_\text{SSH} - {\bm \Sigma}_S(E) - {\bm\Sigma}_D(E)\right]^{-1},
\label{greens}
\end{equation}
where $E$ is the electronic energy and ${\mathbb I}$ denotes the $3N\times 3N$ identity matrix. In Eq.~\ref{greens}, $H_\text{SSH}$ denotes the Hamiltonian of the central system (an extended SSH chain under light irradiation), while ${\bm\Sigma}_S$ and ${\bm\Sigma}_D$ represent the self-energy matrices of the source and drain electrodes, respectively. The effective Hamiltonian is thus given by $H_{\text{eff}} = H_\text{SSH} + {\bm\Sigma}_S + {\bm\Sigma}_D$. The advanced Green's function is obtained as $G^a(E) = \left[G^r(E)\right]^\dagger$.  

The self-energies effectively replace the infinite electrodes with energy-dependent matrices that can be evaluated analytically. In the present model, the source electrode is coupled exclusively to the first site of the central SSH chain, while the drain electrode is coupled exclusively to the last site. Consequently, the full self-energy matrices entering Eq.~\ref{greens} assume a diagonal form with a single non-zero entry each. Specifically,
\begin{eqnarray}
{\bm \Sigma}_S = \text{diag}\left(\Sigma_S, 0, \ldots, 0\right),\\
{\bm \Sigma}_D = \text{diag}\left(0, \ldots, 0,\Sigma_D\right),
\label{sigma-mat}
\end{eqnarray}
where $\Sigma_S$ and $\Sigma_D$ are the complex self-energy functions of the source and drain, respectively.
For a one-dimensional semi-infinite electrode, the self-energy takes the form
\begin{equation}
\Sigma_{S(D)}(E) = \frac{\tau_{S(D)}^2}{2 t_0^2}\left[E - \epsilon_0 - i\sqrt{4t_0^2 - \left(E-\epsilon_0\right)^2}\,\right],
\end{equation}
where $\tau_{S(D)}$ is the coupling between the source (drain) and the central system, $t_0$ is the hopping amplitude within the electrodes, and $\epsilon_0$ is the onsite energy of the electrode sites. The real part $\mathrm{Re}\!\left[\Sigma_{S(D)}\right]$ corresponds to a shift of the device energy levels due to electrode coupling, while the imaginary part $\mathrm{Im}\!\left[\Sigma_{S(D)}\right]$ introduces broadening associated with the finite lifetime of the device states, reflecting the possibility of electron escape into the reservoirs.  

The two-terminal transmission probability is then expressed as~\cite{datta1997electronic,datta2005quantum}
\begin{equation}
T(E) = \mathrm{Tr}\!\left[ \Gamma_S(E) \, G^r(E) \, \Gamma_D(E) \, G^a(E)\right],
\end{equation}
where $\Gamma_S$ and $\Gamma_D$ are the coupling matrices of the source and drain electrodes, respectively. These matrices are obtained from the self-energies as
\begin{equation}
\Gamma_{S(D)}(E) = i\left[\Sigma_{S(D)}(E) - \Sigma_{S(D)}^\dagger(E)\right].
\end{equation}

Finally, the net junction current for a bias voltage $V$ at absolute zero temperature~\cite{datta1997electronic,datta2005quantum} is given by
\begin{equation}
I({\mathcal V}) = \frac{2e}{h} \int_{E_F - \frac{e{\mathcal V}}{2}}^{E_F + \frac{e{\mathcal V}}{2}} T(E) \, dE,
\end{equation}
where $E_F$ is the Fermi energy, and $T(E)$ is the two-terminal transmission probability, evaluated using the Green's function formalism.

\section{\label{sec3}Results and Discussion}
First, we outline the various parameters and their units used in this study. Energy values are given in electron-volt (eV). The number of unit cells considered is $N=7$, that is the chain is made up of 21 sites. For simplicity, the voltage-independent on-site potential \(\epsilon_{n,\alpha}^0\) and the on-site potential for the electrodes \(\epsilon_0\) are both set to zero. The voltage-dependent on-site potential \(\epsilon_n({\mathcal V})\) varies linearly across the molecule's length within the range \([-{\mathcal V}/2:{\mathcal V}/2]\). The coupling strengths between the electrodes and the chain, $\tau_S$ and $\tau_D$, are set to $0.75\,$eV. The hopping amplitude within the electrodes is fixed at $t_0 = 2.5\,$eV, while for the SSH chain we take $t_1 = t_2 = t_3 = t = 1\,$eV in the absence of light. These choices of tight-binding parameters ensure that the system operates within the wide-band limit~\cite{wide-band}. It is important to mention that our numerical calculations use a specific set of parameter values that align with other theoretical studies in the literature. Alternative parameter sets can be considered, but our main findings and the physical interpretations remain consistent, as verified through detailed numerical analysis. The vector potential is given in units of \(ea/\hbar\), where $a$ is the lattice constant and considered 1$\,$\AA~for simplicity.

Due to the time-periodicity of the driving field, a periodically driven \(\mathbb{D}\)-dimensional lattice can be mapped onto an equivalent undriven \((\mathbb{D}+1)\)-dimensional lattice~\cite{PhysRevLett.110.200403,sambe1973steady,grifoni1998driven}. In such systems, the original Bloch bands of the undriven lattice evolve into a set of Floquet-Bloch bands, with the inter-band coupling determined by the driving frequency. This effective time-independent \((\mathbb{D}+1)\)-dimensional system can be visualized as the original system connected to multiple virtual replicas stacked along the additional (Floquet) dimension. In the high-frequency limit, the Floquet bands decouple, and the dynamics are primarily governed by the zeroth-order Floquet-Bloch band (\(p = q = 0\)) in Eq.~\ref{effhop}, while the contributions from higher-order terms in \(p\) and \(q\) become negligibly small. As a result, the coupling between the original system and its virtual copies becomes effectively suppressed. However, this picture breaks down in the low-frequency regime, where strong coupling between the parent system and its Floquet replicas emerges. In this regime, multiple virtual copies actively contribute to the dynamics, effectively increasing the system's dimensionality and spatial extent. This enlarged system size can lead to a reduction in the phase-coherence (or phase-relaxation) length, which is further diminished at finite temperatures. Consequently, achieving favorable transport properties becomes increasingly challenging in the low-frequency regime. A justification for considering only the zeroth-order Floquet-Bloch band (\(p = q = 0\)) in the high-frequency regime is provided in Appendix~\ref{appc}.

Given the considerations discussed above, we confine our analysis to the high-frequency regime, where \(\hbar \Omega > 4\sqrt{\frac{t_1^2+t_2^2+t_3^2}{3}}\) (The band width for the system can be estimated from the analysis given in Appendix~\ref{appb}). For our calculations, we fix $\hbar \Omega = 6\,$eV, corresponding to a driving frequency of $\sim 10^{15}\,\text{Hz}$, i.e., in the near-ultraviolet (UV) or extreme ultraviolet (XUV) range. Such frequencies are experimentally accessible via ultrashort femtosecond laser pulses~\cite{uv1,uv2}. The corresponding electric and magnetic fields are approximately $E_0 \sim 6\,$V/\AA\ and $B_0 \sim 200\,$T, with an intensity of $I \sim 10^{14}\,\text{W/cm}^2$. Although these values are large, they serve only as illustrative estimates within our model framework. Heating effects may arise in the sample, however, they can be avoided by employing artificial or engineered systems~\cite{PhysRevB.88.245422,artificial}, which effectively enlarge the lattice constant and allow lower frequencies and reduced field intensities. For instance, increasing the lattice spacing to $\sim 10\,$\AA, reduces the fields to $E_0 \sim 0.06\,$V/\AA, $B_0 \sim 2\,$T, and $I \sim 10^{10}\,\text{W/cm}^2$. Most importantly, since the systems under consideration are non-magnetic, the magnetic component of the electromagnetic field does not significantly modify their electronic structure. Furthermore, much higher light intensities have been reported in recent experiments~\cite{cwd1,cwd2}, confirming the feasibility of our chosen parameter regime without the risk of sample damage. With this high-frequency framework established, we proceed to discuss our results, highlighting the influence of various system parameters.

\begin{figure}[ht]
\includegraphics[width=0.2375\textwidth]{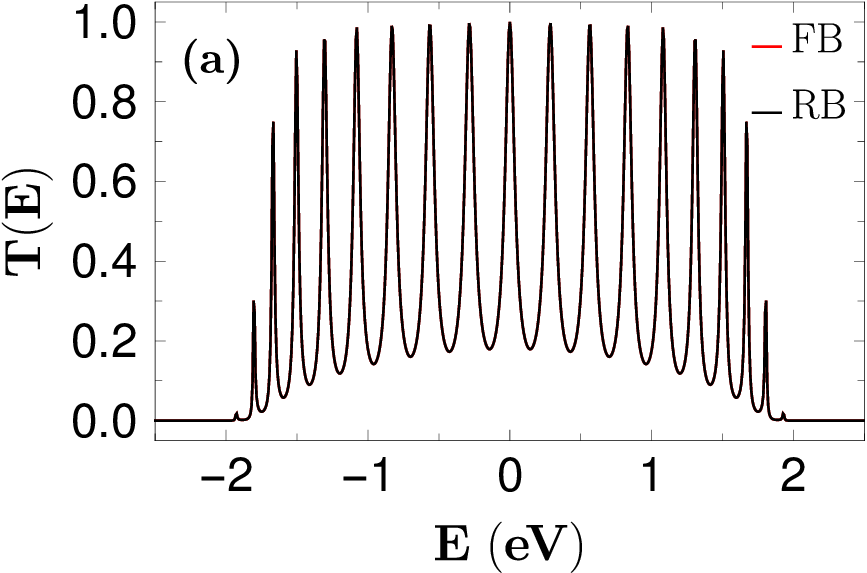}
\includegraphics[width=0.2375\textwidth]{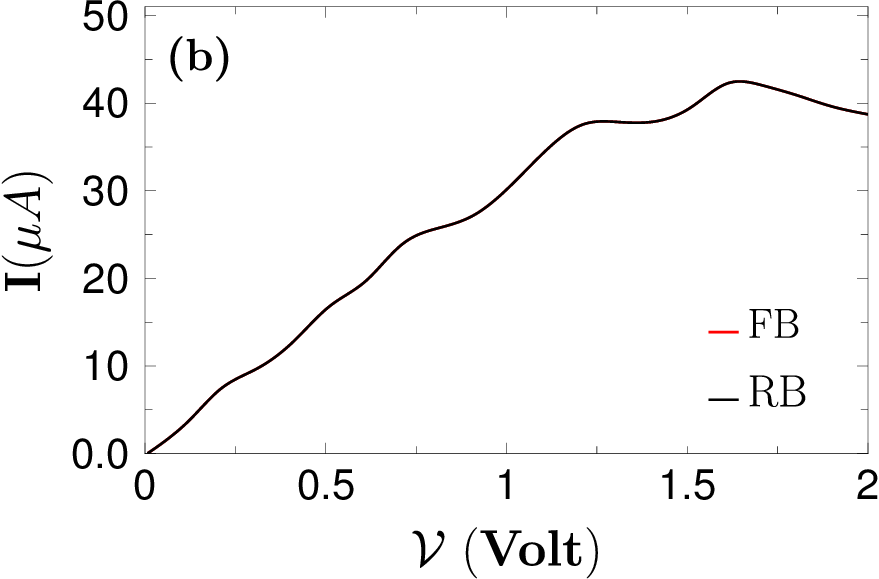}
\includegraphics[width=0.2375\textwidth]{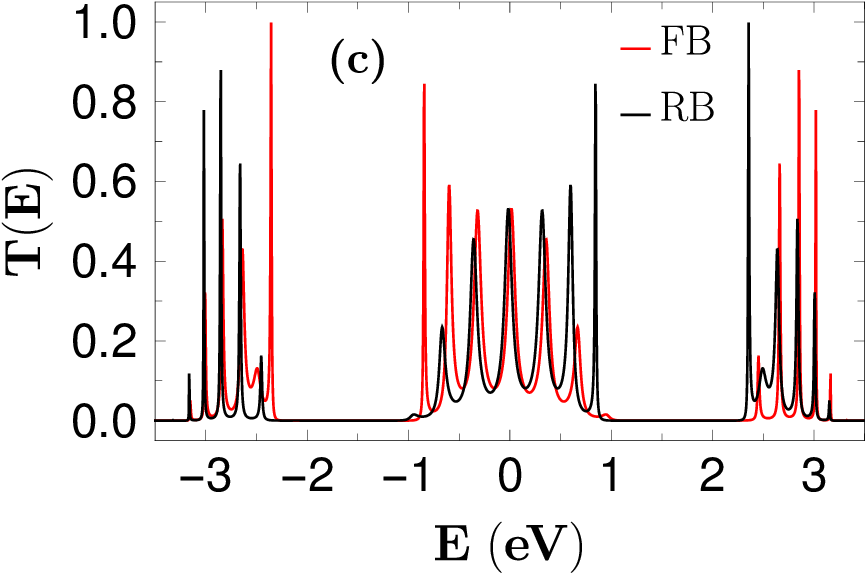}
\includegraphics[width=0.2375\textwidth]{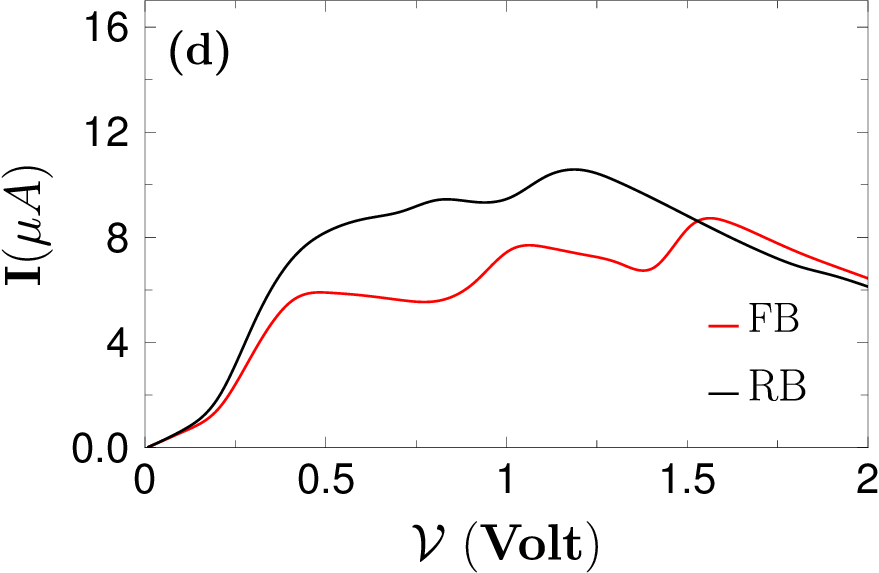}
\caption{(Color online.) Top panel: Isotropic hopping configuration. Bottom panel: Anisotropic hopping configuration. (a), (c) Transmission probability as a function of energy at a bias ${\mathcal V}=0.5\,$Volt. (b), (d) Current-voltage characteristics. The number of unit cells is $N=7$. The hopping amplitudes for the isotropic case are fixed as $t_1=t_2=t_3=1\,$eV. For the anisotropic case $t_1 = 1\,$eV, $t_2 = 1.5\,$eV, and $t_3 = 1.25\,$eV. For the computation of the current, Fermi energy is fixed at $E_F =0.5\,$eV. Red and black colors denote the results for the forward and reverse bias conditions, respectively.}
\label{Fig2}
\end{figure}

We begin our investigation with the simplest configuration of the extended SSH chain, where the hopping integrals are \textit{isotropic}, i.e., $t_1 = t_2 = t_3 = 1\,$eV, and the system is not exposed to light irradiation. In Fig.~\ref{Fig2}(a), we present the two-terminal transmission probability as a function of electron energy. The transmission spectra for \textit{forward bias} and \textit{reverse bias} are shown in red and black, 
respectively. Owing to the uniform hopping parameters, the system retains its \textit{centrosymmetry}. Consequently, the transmission spectra are identical under both bias directions, as the red and black curves perfectly overlap, indicating the \textit{absence of rectification}. This behavior is corroborated in Fig.~\ref{Fig2}(b), where the current-voltage ($I-V$) characteristics demonstrate complete symmetry between forward and reverse bias conditions. In Fig.~\ref{Fig2}(c), we consider a more general case with \textit{anisotropic hopping integrals}, specifically $t_1 = 1\,$eV, $t_2 = 1.5\,$eV, and $t_3 = 1.25\,$eV. The asymmetry in hopping strengths breaks centrosymmetry of the system, resulting in distinct transmission spectra for forward (red) and reverse (black) biases. The breaking of this symmetry can be understood from the relation $\theta_I H_{FB} \theta_I^{-1} \neq H_{RB}$, where $H_{FB}$ and $H_{RB}$ denote the Hamiltonians under forward and reverse biases, respectively, and $$\theta_I = \begin{pmatrix}
0 & 0 & 1\\
0 & 1 & 0\\
1 & 0 & 0
\end{pmatrix}$$ represents the inversion operator for a one unit cell. This directional dependence in transport reveals the emergence of \textit{rectification behavior}. Figure~\ref{Fig2}(d) confirms this finding, showing that at a fixed Fermi energy of $E_F = 0.5\,$eV, the $I-V$ characteristics exhibit a clear difference between forward and reverse bias currents, indicating a \textit{non-zero rectification ratio} in the proposed setup.

\begin{figure}[ht]
\includegraphics[width=0.4\textwidth]{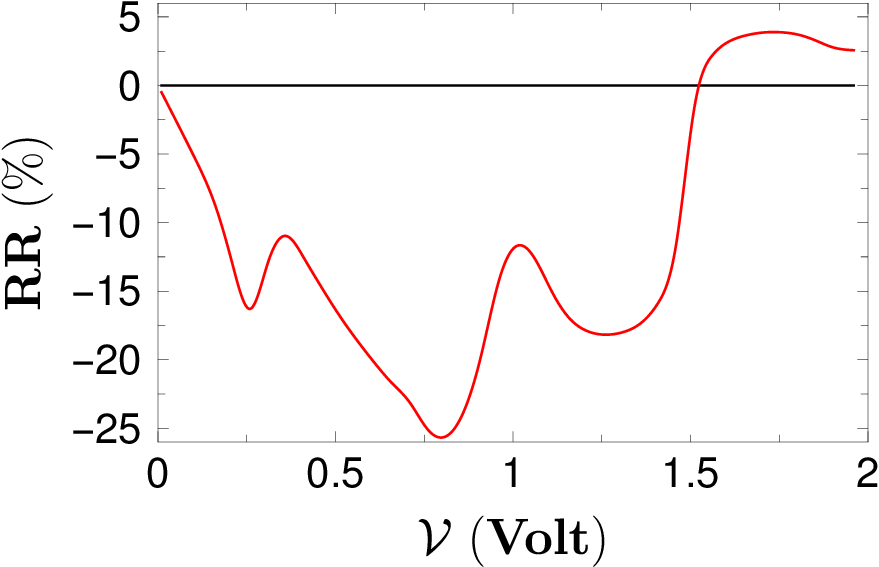}
\caption{(Color online.) Rectification ratio $RR$ as a function of voltage $V$. Black and red colors denote the results for the isotropic and anisotropic cases. The hopping amplitudes for the isotropic case are fixed as $t_1=t_2=t_3=1\,$eV. For the anisotropic case $t_1 = 1\,$eV, $t_2 = 1.5\,$eV, and $t_3 = 1.25\,$eV. The Fermi energy is fixed at $E_F =0.5\,$eV.}
\label{Fig3}
\end{figure}

To gain deeper insight and a quantitative perspective on the rectification behavior, we compute the rectification ratio for the previously discussed parameter sets and plot it as a function of applied voltage in Fig.~\ref{Fig3}. For the case of isotropic hopping integrals, $t_1 = t_2 = t_3 = 1\,$eV, the rectification ratio remains zero across the entire voltage range, as indicated by the flat black line in Fig.~\ref{Fig3}. This result is consistent with the identical transmission spectra under forward and reverse bias conditions, which preserve the centrosymmetry of the system. 
In contrast, for the anisotropic case with $t_1 = 1\,$eV, $t_2 = 1.5\,$eV, and $t_3 = 1.25\,$eV, the rectification ratio shows a clear dependence on the applied voltage. As shown in Fig.~\ref{Fig3}, the ratio varies significantly with bias, reaching a maximum value of approximately $25\%$. This behavior reflects the asymmetric transport properties induced by the broken centrosymmetry and confirms the emergence of rectification in the system.

We now have established the fact that asymmetric hopping amplitudes break the centrosymmetry of the extended SSH chain, which makes rectification possible. In this work, we suggest a new way to introduce such asymmetry. The method is to irradiate the chain with light. Theoretical analysis shows that irradiation modifies the 
\begin{figure}[ht]
\includegraphics[width=0.2375\textwidth]{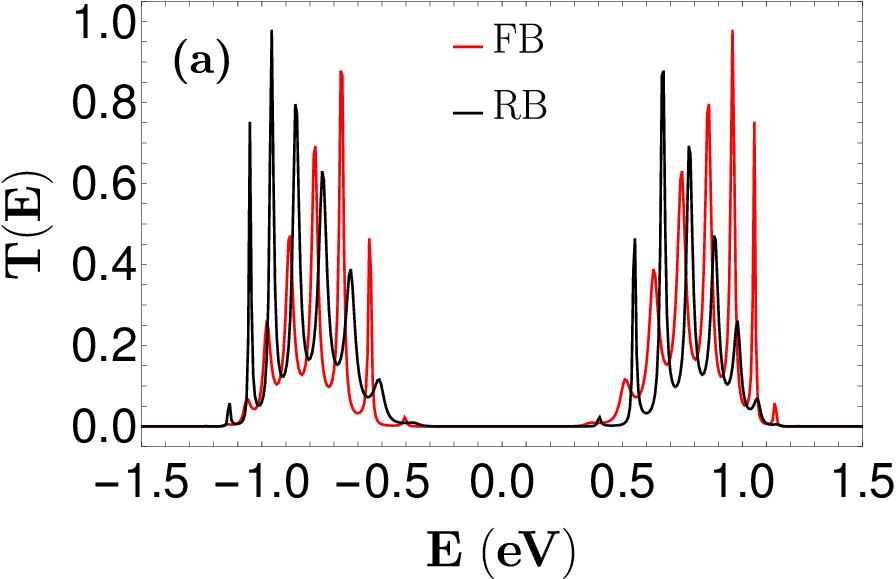}
\includegraphics[width=0.2375\textwidth]{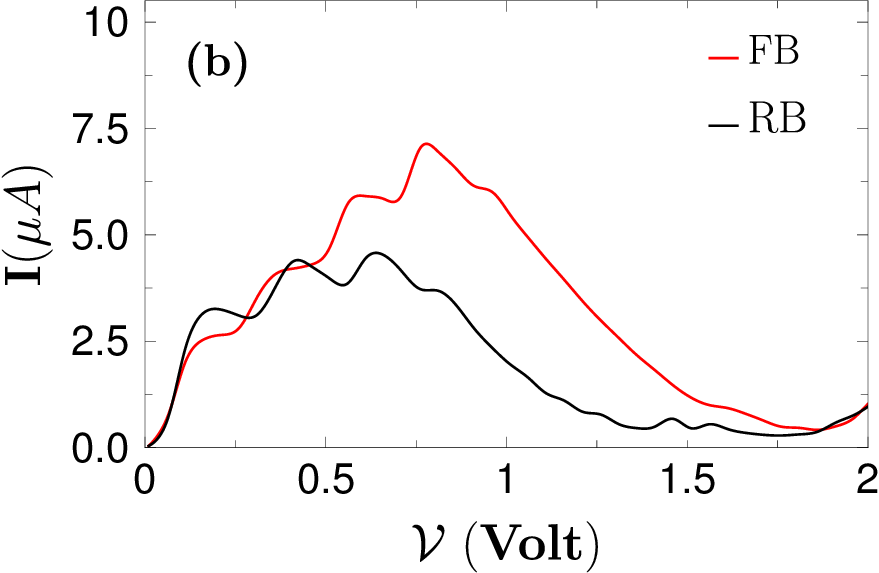}
\caption{(Color online.) (a) Transmission probability as a function of energy and (b) current as a function of voltage in the presence of light. The number of unit cells of the chain $N=7$. The hopping amplitudes are taken to be identical that is $t_1 = t_2 = t_3 = 1\,$eV. The Fermi energy is fixed at $E_F=0.5\,$eV. The light parameters are $A_x = 2.5$, $A_y = 0.5$, and $\phi = 0$. Red and black colors denote the results for the forward and reverse bias conditions, respectively.}
\label{Fig4}
\end{figure}
hopping amplitudes depending on their direction, which leads to the breaking of centrosymmetry even when all three hopping strengths are equal. This feature offers a clear advantage over the dimerized SSH chain, where symmetry cannot be broken by light if the electrodes are symmetrically connected. Moreover, light not only induces asymmetry but also provides additional control over the rectification behavior through tunable parameters such as amplitude and phase. Therefore, in Fig.~\ref{Fig4}(a), we present the two-terminal transmission probability as a function of electron energy under both forward and reverse bias conditions, in the presence of light. The light parameters are set to $A_x = 2.5$, $A_y = 0.5$, and phase $\phi = 0$. The hopping amplitudes are taken identical that is $t_1=t_2=t_3=1\,$eV. The number of unit cell is fixed as $N=7$. The application of light modulates the effective hopping amplitudes and breaks the spatial symmetry of the system as is evident from the different transmission spectrum under two opposite biases. we also note that the allowed energy window within which non-zero transmission probability is there, is reduced compared to the irradiation-free case. This reduction in the energy window is according to the Eq.~\ref{effhop1}. Figure~\ref{Fig4}(b) displays the $I-V$ characteristics of the system under light irradiation, using the same light parameters as in Fig.~\ref{Fig4}(a), keeping the Fermi energy fixed at $E_F=0.5\,$eV. Within the chosen voltage window (0-2~V), the forward bias current is significantly larger than the reverse bias current, leading to pronounced asymmetry and resulting in high rectification values.

These results clearly demonstrate that light irradiation not only breaks spatial symmetry but also serves as an effective external means to tune current rectification in the extended SSH chain. It is important to note that in Fig.~\ref{Fig4}, we analyzed a special case corresponding to isotropic hopping ($t_1 = t_2 = t_3$) with an odd number of unit cells ($N = 7$). In this scenario, we find that $\theta_I H_{FB} \theta_I^{-1} = H_{RB}$ in the absence of light, whereas $\theta_I H_{FB} \theta_I^{-1} \neq H_{RB}$ once light is introduced. Notably, this irradiation-induced inversion symmetry breaking occurs only when (i) the number of unit cells is odd, and (ii) $A_x, A_y \neq 0$, with $\phi$ taking any value except for circular polarization ($A_x = A_y$, $\phi = \pi/2$). Conversely, if the number of unit cells is even, or if either $A_x = 0$ or $A_y = 0$, or if the light is circularly polarized, the system preserves inversion symmetry under opposite bias even in the presence of light.

An interesting feature we observe in the transmission plots of Figs.~\ref{Fig2}(c) and \ref{Fig4}(a) is that the transmission functions are symmetric under $E \rightarrow -E$ and FB$\rightarrow$RB. This apparent symmetry arises from a generalized particle-hole-like symmetry of the system. For the three-sublattice chain studied here, a generalized particle-hole-like symmetry can be defined via a unitary operator $\theta$, which flips the sign of specific sublattices. In the basis $\left(A_1,B_1,C_1,A_2,B_2,C_2,\ldots\right)$, this operator can be written as $\theta = \text{diag}(1,-1,1,1,-1,1,\ldots)$, where the entries correspond to sublattices $A$, $B$, and $C$ in each unit cell. To be consistent with the standard definition of particle-hole (PH) symmetry, we note that PH symmetry is inherently anti-unitary and is represented by the operator
$
\mathcal{T} = \theta \mathcal{K},
$
where $\mathcal{K}$ denotes complex conjugation. Accordingly, the generalized PH-like symmetry relating the forward- and reverse-bias Hamiltonians takes the form
$
\theta H_{FB}^* \theta^{-1} = - H_{RB}.
$
This is the real-space analogue of the conventional PH condition $U H^* U^{-1} = -H$ (or equivalently $U H({\mathbf k})^*U^{-1} = -H(-{\mathbf k})$ in momentum space, where $U$ is an unitary matrix and ``$*$" denotes complex conjugation). Although the onsite potentials break exact particle-hole symmetry, this generalized anti-unitary symmetry nevertheless relates the forward- and reverse-bias configurations. As a result, for each eigenstate of $H_{FB}$ with energy $E$, there exists a corresponding eigenstate of $H_{RB}$ with energy $-E$, naturally explaining the apparent symmetry in the transmission plots of Figs.~2(c) and 4(a) under $E \rightarrow -E$ and FB$\rightarrow$RB.

\begin{figure}[ht]
\includegraphics[width=0.4\textwidth]{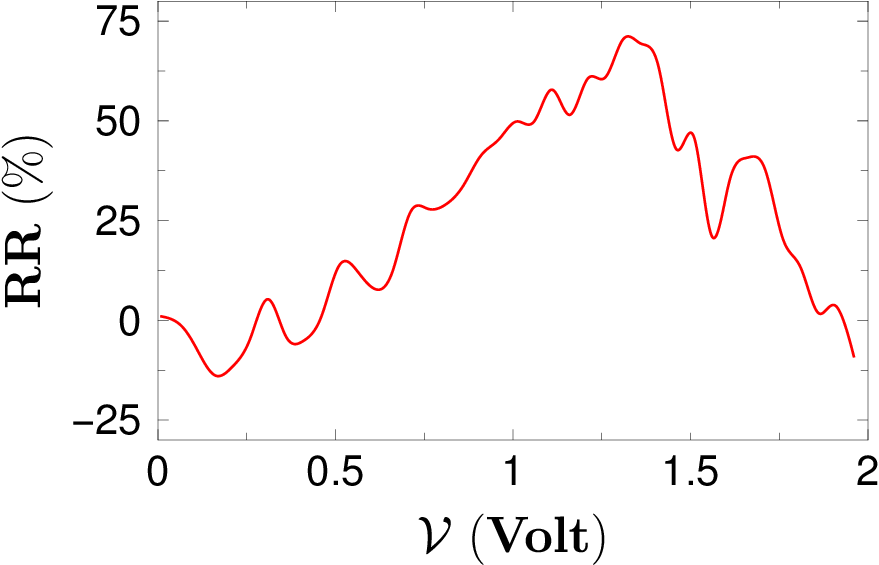}
\caption{(Color online.) Rectification ratio as a function of voltage $V$ in the presence of light. All the systems parameters are identical to Fig.~\ref{Fig4} that is the number of unit cells of the chain $N=7$. The hopping amplitudes are $t_1 = t_2 = t_3 = 1\,$eV. The Fermi energy is fixed at $E_F=0.5\,$eV. The light parameters are $A_x = 2.5$, $A_y = 0.5$, and $\phi = 0$.}
\label{Fig5}
\end{figure}
To gain quantitative insight into the rectification behavior under the chosen light parameters, we compute and plot the rectification ratio as a function 
of the applied voltage, as shown in Fig.~\ref{Fig5}. In particular, the rectification ratio reaches a peak value of approximately $75\%$ at a bias voltage of around $1.3\,$V. A positive value of $RR$ indicates that the forward bias current exceeds its reverse bias counterpart.

To identify the conditions for optimal current rectification, we systematically explore the parameter space introduced by light irradiation, focusing on the light amplitudes $A_x$, $A_y$, 
\begin{figure}[ht!]
\includegraphics[width=0.23\textwidth]{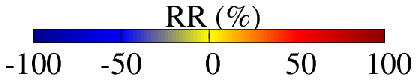}\\
\includegraphics[width=0.238\textwidth]{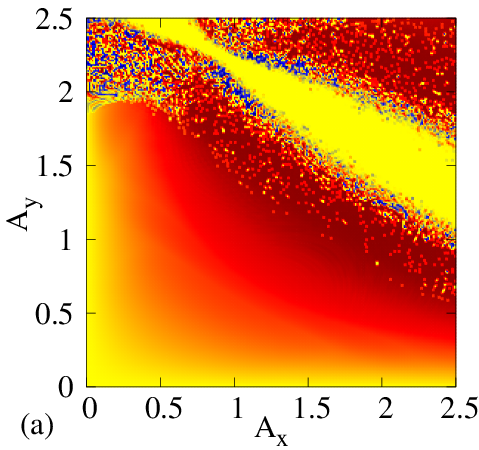}
\hfill
\includegraphics[width=0.238\textwidth]{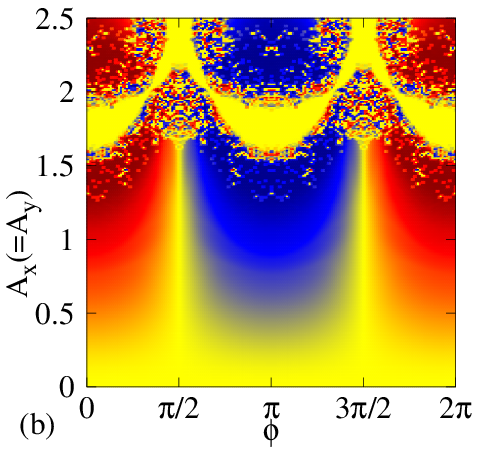}
\caption{(Color online.) Color density plot for maximum $RR$ as functions of (a) $A_x$ and $A_y$ with $\phi = 0$, (b) $A_x\left(=A_y\right)$ and $\phi$. The color bar denotes the value of maximum $RR$. Dark red and dark blue regions denote a positive and negative high degree of rectifications, respectively, while the yellowish color represents zero $RR$. The hopping amplitudes are $t_1 = t_2 = t_3 = 1\,$eV. The Fermi energy is fixed at $E_F=0.5\,$eV. The definition of maximum $RR$ is given in the texts.}
\label{Fig6}
\end{figure}
and the polarization phase $\phi$. For this purpose, we present density plots of the maximum rectification efficiency, evaluated over the voltage range $0-2\,$V, as functions of these light parameters. The color code for the density plots is as follows. Red regions correspond to high $RR$ dominated by forward bias, blue regions indicate high $RR$ dominated by reverse bias, and yellow regions represent poor rectification performance. All physical parameters of the system are kept the same as in Fig.~\ref{Fig4}, 
and the Fermi energy is fixed at $E_F = 0.5\,$eV. In Fig.~\ref{Fig6}(a), the density plot shows the maximum $RR$ as functions of $A_x$ and $A_y$, with the phase fixed at $\phi = 0$. The results reveal that rectification remains significantly high over a wide region of the parameter space, with efficiency reaching values as high as $90\%$. In Fig.~\ref{Fig6}(b), we present the maximum rectification efficiency as functions of $A_x (= A_y)$ and $\phi$. Here also we observe high rectification ratio over a wide range of parameter space. The maximum $RR$ exceeds $90\%$, dominated by both the forward and reverse bias conditions. The plot is symmetric about the $\phi = \pi$ line, which arises due to the following reason. The origin of this behavior lies in the two angular bonds of our system (see Fig.~1). Their hopping amplitudes are renormalized under irradiation, as described by Eq.~\ref{gam}, and explicitly depend on the phase $\phi$. Under the transformation $\phi \rightarrow \phi + \pi$, the effective hopping amplitudes of the two angular bonds are interchanged. However, since the Hamiltonian remains unchanged under this swap, the rectification ratio exhibits symmetry with respect to $\phi=\pi$. These findings clearly demonstrate that the rectification ratio can be effectively tuned by varying the field amplitudes and the phase of the incident light, making this setup a promising platform for light-controlled quantum transport.

\begin{figure}[ht!]
\includegraphics[width=0.238\textwidth]{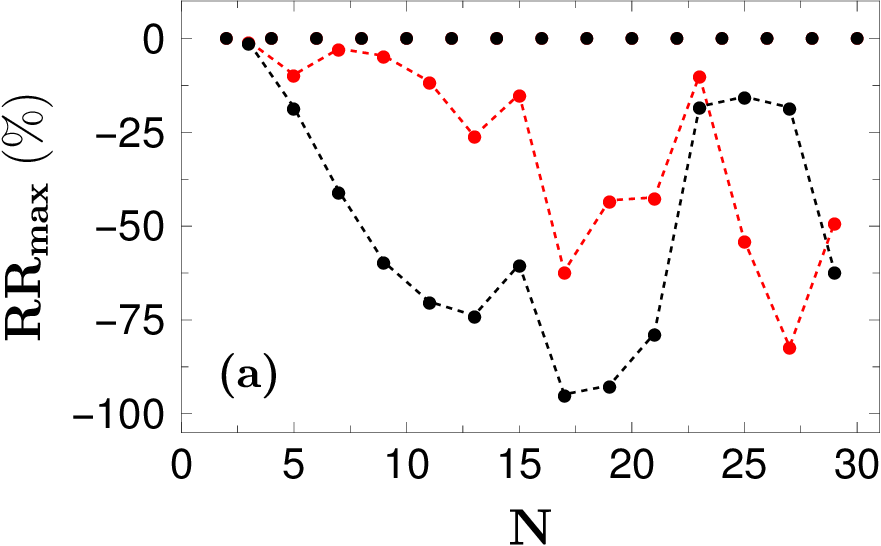}
\hfill
\includegraphics[width=0.238\textwidth]{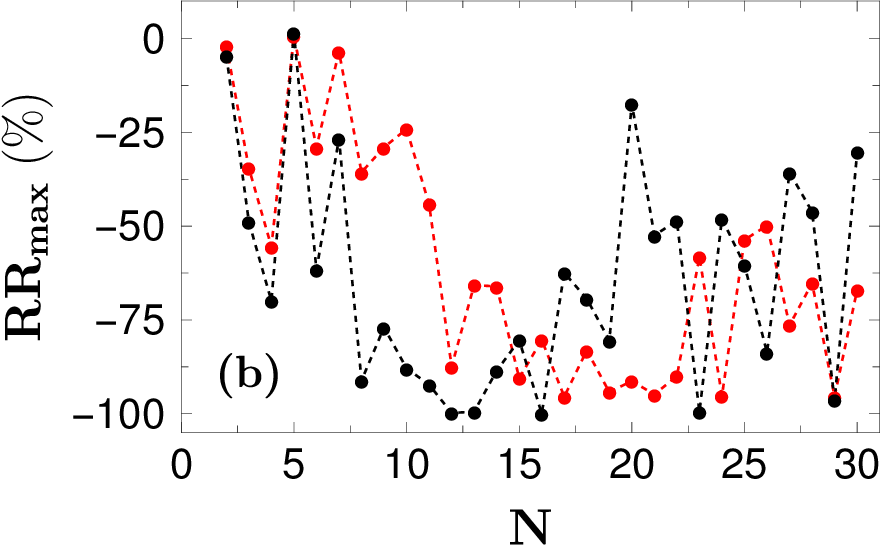}
\caption{(Color online.) Maximum $RR$ as function of number of unit cells $N$. (a) Isotropic case with $t_1=t_2=t_3=1\,$eV and (b) anaisotropic case with $t_1=1\,$eV, $t_2=1.25\,$eV, and $t_3=1.5\,$eV. Two sets of light parameters are considered, namely $A_x=2.5$, $A_y=0.5$, $\phi=0$ (black), and $A_x=1$, $A_y=1$, $\phi = 0$ (red) color. The Fermi energy is fixed at $E_F=0.5\,$eV. The definition of maximum $RR$ is the same as in Fig.~\ref{Fig6}.}
\label{Fig7}
\end{figure}

Finally, we study the rectification performance as a function of the system dimension. To this end, we plot the maximum rectification ratio, as defined in Fig.~\ref{Fig6}, as a function of the number of unit cells (\(N\)) in Fig.~\ref{Fig7}. Two different light parameters are considered for this purpose, namely \(A_x = 2.5\), \(A_y = 0.5\), and \(\phi = 0\) (black), and \(A_x = 1\), \(A_y = 1\), and \(\phi = 0\) (red). In Fig.~\ref{Fig7}(a), we present the isotropic hopping case with \(t_1 = t_2 = t_3 = 1\,\text{eV}\), while Fig.~\ref{Fig7}(b) corresponds to the anisotropic hopping scenario with \(t_1 = 1\,\text{eV}\), \(t_2 = 1.25\,\text{eV}\), and \(t_3 = 1.5\,\text{eV}\). In the isotropic case, only systems with an odd number of unit cells exhibit finite rectification, whereas even-numbered systems show identically zero \(RR\). As discussed earlier, 
in the isotropic configuration, the system retains inversion (centrosymmetric) symmetry for even \(N\), resulting in identical forward and reverse bias currents and hence zero rectification. In contrast, for odd \(N\), this symmetry is broken, leading to finite \(RR\). Among the two light configurations, the anisotropic illumination (\(A_x \neq A_y\), black curve) produces a stronger rectification response than the symmetric illumination (\(A_x = A_y\), red curve). For the black curve, \(|RR|\) reaches nearly \(100\%\) for \(N = 17\) and \(19\), while for the red curve it attains about \(80\%\) around \(N = 27\). For the anisotropic hopping case, shown in Fig.~\ref{Fig7}(b), finite rectification appears for both even and odd \(N\), as the intrinsic hopping asymmetry (\(t_1 \neq t_2 \neq t_3\)) breaks inversion symmetry irrespective of system length. The rectification ratio exhibits pronounced oscillations with \(N\), which are more irregular and of larger amplitude compared to the isotropic case. These oscillations arise from the combined effects of structural and light-induced asymmetries. Overall, the anisotropic case demonstrates more favorable rectification performance, with both light configurations yielding \(|RR|\) values in the range of \(80{-}100\%\) over a broad range of \(N\). In all cases, \(RR\) remains negative for the chosen light parameters, indicating that the reverse bias current dominates over the forward bias current.

\section{\label{con}Conclusion}

In conclusion, we have demonstrated rectification behavior in an extended SSH chain with a trimerized bond configuration arranged in a zigzag pattern under light irradiation. The central idea is to induce spatial anisotropy in the hopping amplitudes through the irradiation effect. While the extended SSH chain with anisotropic hoppings intrinsically breaks inversion symmetry and can independently yield rectification, the irradiation can also generate a similar effect even in the isotropic case, depending on the light polarization. Hence, light serves as an additional control knob to tune and enhance the rectification performance.

Our results show that the forward and reverse bias currents differ significantly in the presence of light, leading to a pronounced rectification effect. The degree of rectification remains highly favorable across a broad range of light parameters. The strong rectification efficiency and optical tunability revealed in this work highlight the potential of light-driven systems as promising candidates for next-generation nanoelectronic rectifiers.

\appendix 

\section{\label{appa}Incorporation of light irradiation}
The influence of light irradiation can be incorporated into the extended SSH chain following Refs.~\cite{PhysRevLett.110.200403,PhysRevB.88.245422}. We consider the situation where the incident light is directed perpendicular to the plane of the chain. Once the system is exposed to light, its Hamiltonian acquires an explicit time dependence. Owing to the periodic nature of the driving, the Hamiltonian can generally be assumed to be periodic in both space and time, namely  
\[
H(\vec{x} + \vec{a}, \tau + \mathbb{T}) = H(\vec{x}, \tau + \mathbb{T}) = H(\vec{x} + \vec{a}, \tau),
\]  
where $\vec{a}$ is the lattice translation vector, $\mathbb{T}$ is the driving period, and $\mathbb{T} = 2\pi/\Omega$ with $\Omega$ denoting the frequency of irradiation. Under such conditions, the Floquet--Bloch ansatz provides the eigenstates in the form  
\begin{equation}
\lvert\Psi_{\alpha,\mathbf{k}}(\mathbf{x},\tau)\rangle = \mathrm{e}^{(i\mathbf{k}\cdot\mathbf{x} - i\epsilon_{\alpha,\mathbf{k}}t)} \lvert u_{\alpha,\mathbf{k}}(\mathbf{x},\tau)\rangle,
\end{equation}  
where $\epsilon_{\alpha,\mathbf{k}}$ is the quasi-energy associated with the $\alpha$th Floquet-Bloch mode $\lvert u_{\alpha,\mathbf{k}}(\mathbf{x},\tau)\rangle$, and $\mathbf{k}$ is the crystal momentum. The Floquet--Bloch states are periodic in both $\mathbf{x}$ and $\tau$, and they span a Hilbert space constructed as the direct product between the space of $\mathbb{T}$-periodic functions and the ordinary Hilbert space $\mathcal{H}$, i.e., $\mathcal{S} = \mathcal{H}\otimes\mathbb{T}$. This extended space is usually referred to as the Sambe space~\cite{sambe1973steady}.  

To proceed, we begin with a general tight-binding Hamiltonian for a one-dimensional chain, expressed as  
\begin{equation}
H = \sum_\alpha \sum_{n,m} t_{nm} c^\dagger_{\alpha,n}(\tau) c_{\alpha,m}(\tau),
\label{ham-ex}
\end{equation}  
where $c^\dagger_{\alpha,n}(\tau)$ and $c_{\alpha,n}(\tau)$ represent the time-dependent fermionic creation and annihilation operators, respectively, and $t_{nm}$ denotes the hopping amplitude between sites $n$ and $m$. The Fourier transform of these operators is introduced as  
\begin{eqnarray}
c_{\alpha,\mathbf{k}}(\tau) &=& \frac{1}{\sqrt{N}}\sum_{m} c_{\alpha,\mathbf{k}}(\tau)\mathrm{e}^{i\mathbf{k}\cdot\mathbf{R}_m}\\
c_{\alpha,\mathbf{k}}^\dagger(\tau) &=& \frac{1}{\sqrt{N}}\sum_{m} c_{\alpha,\mathbf{k}}^\dagger(\tau)\mathrm{e}^{-i\mathbf{k}\cdot\mathbf{R}_m}.
\end{eqnarray}  

The inverse Fourier transforms are accordingly  
\begin{eqnarray}
c_{\alpha,m}(\tau) &=& \frac{1}{\sqrt{N}}\sum_{\mathbf{k}} c_{\alpha,\mathbf{k}}(\tau)\mathrm{e}^{-i\mathbf{k}\cdot\mathbf{R}_m}\\
c_{\alpha,m}^\dagger(\tau) &=& \frac{1}{\sqrt{N}}\sum_{\mathbf{k}} c_{\alpha,\mathbf{k}}^\dagger(\tau)\mathrm{e}^{i\mathbf{k}\cdot\mathbf{R}_m}.
\end{eqnarray}  

Using these definitions, Eq.~\ref{ham-ex} can be rewritten as  
\begin{equation}
H = \sum_{\alpha,\mathbf{k}} \sum_{n,m} t_{nm} c^\dagger_{\alpha,\mathbf{k}}(\tau) c_{\alpha,\mathbf{k}}(\tau)\mathrm{e}^{i\mathbf{k}\cdot\left(\mathbf{R}_n - \mathbf{R}_m\right)}.
\end{equation}  
The periodicity in time allows the operators to be expanded as  
\begin{eqnarray}
c_{\alpha,\mathbf{k}}(\tau) &=& \sum_p c_{\alpha,\mathbf{k},p}\mathrm{e}^{ip\Omega\tau}\\
c^\dagger_{\alpha,\mathbf{k}}(\tau) &=& \sum_p c^\dagger_{\alpha,\mathbf{k},p}\mathrm{e}^{-ip\Omega\tau}.
\end{eqnarray}  

Substituting these expansions, the Hamiltonian becomes  
\begin{eqnarray}
H_\mathbf{k} &=& \sum_{\alpha,\mathbf{k}} \sum_{n,m} \sum_{p,q} t_{nm} \mathrm{e}^{i\mathbf{k}\cdot\left(\mathbf{R}_n - \mathbf{R}_m\right)} \mathrm{e}^{-i\Omega\tau(p-q)} c^\dagger_{\alpha,\mathbf{k},p} c_{\alpha,\mathbf{k},q}\nonumber\\
&=&\sum_{\alpha,\mathbf{k}} \sum_{n,m} \sum_{p,q} \tilde{t}_{nm} \lvert U_{\alpha,\mathbf{k},p}\rangle \mathrm{e}^{-i\Omega\tau(p-q)} \langle U_{\alpha,\mathbf{k},q}\rvert,
\end{eqnarray}  
with $\tilde{t}_{nm} = t_{nm}\mathrm{e}^{i\mathbf{k}\cdot\left(\mathbf{R}_n - \mathbf{R}_m\right)}$ representing the momentum-modified hopping parameter.  

Diagonalizing $H_\mathbf{k}$ within Sambe space provides the quasi-energies as  
\begin{eqnarray}
\epsilon_{\alpha,\mathbf{k}} &=& \langle\langle U_{\alpha,\mathbf{k},p} \lvert\mathcal{H}_\mathbf{k}\rvert U_{\alpha,\mathbf{k},p}\rangle\rangle\nonumber\\
&=&\frac{1}{\mathbb{T}}\int_0^\mathbb{T} \langle U_{\alpha,\mathbf{k},p} \lvert\mathcal{H}_\mathbf{k}\rvert U_{\alpha,\mathbf{k},p}\rangle \mathrm{d}\tau,
\end{eqnarray}  
where the Floquet Hamiltonian is defined as $\mathcal{H}_\mathbf{k} = H_\mathbf{k} - i\hbar\frac{\partial}{\partial\tau}$. Evaluating the scalar product leads to  
\begin{eqnarray}
\epsilon_{\alpha,\mathbf{k}} &=& \sum_{n,m}\frac{1}{\mathbb{T}}\int_0^\mathbb{T} \tilde{t}_{nm}\mathrm{e}^{-i\Omega\tau(p-q)} \mathrm{d}\tau + q\hbar\Omega\delta_{p,q}\nonumber\\
&=&\sum_{n,m} \tilde{t}_{nm}^{p,q} + q\hbar\Omega\delta_{p,q},
\label{eff-ham}
\end{eqnarray}  
where $q\hbar\Omega\delta_{p,q}$ corresponds to the Fourier representation of the derivative operator, and  
\begin{equation}
\tilde{t}_{nm}^{p,q} = \sum_{n,m}\frac{1}{\mathbb{T}}\int_0^\mathbb{T} \tilde{t}_{nm}\mathrm{e}^{-i\Omega\tau(p-q)} \mathrm{d}\tau.
\end{equation}  

Equation~\ref{eff-ham} thus represents an effective static Hamiltonian in which $q\hbar\Omega\delta_{p,q}$ plays the role of on-site energies, and $\tilde{t}_{nm}^{p,q}$ denotes the renormalized hopping elements in the irradiated system. Using the minimal coupling prescription under the dipole approximation~\cite{PhysRevLett.110.200403}, the wavevector transforms as $\mathbf{k}\Rightarrow \mathbf{K}(\tau) = \mathbf{k} + q\mathbf{A}(\tau)/\hbar$, where $q$ is the electronic charge and $\mathbf{A}(\tau)$ is the vector potential of the external light field. The effective hopping integral thereby becomes  
\begin{equation}
\tilde{t}_{nm}^{p,q} = \frac{1}{\mathbb{T}}\int_0^\mathbb{T} t_{nm} \mathrm{e}^{i\mathbf{A}\cdot\left(\mathbf{R}_n - \mathbf{R}_m\right)} \mathrm{e}^{-i\Omega\tau(p-q)}\mathrm{d}\tau.
\label{eff-hop}
\end{equation}  

Here, $\mathbf{R}_{n}$ and $\mathbf{R}_m$ denote the position vectors of the $n$th and $m$th sites, while their difference $\mathbf{R}_{n} - \mathbf{R}_{m}$ corresponds to the bond vector. For a general polarization in the $x$--$y$ plane, the vector potential can be written as  
\begin{equation}
\mathbf{A}(\tau) = {\mathcal A}_x \sin{(\Omega \tau)} \boldsymbol{\hat{x}} + {\mathcal A}_y \sin{(\Omega \tau + \phi)} \boldsymbol{\hat{y}},
\end{equation}  
with ${\mathcal A}_x$ and ${\mathcal A}_y$ being its $x$ and $y$ components, and $\Omega$ the driving frequency. The irradiation is assumed to propagate along the $z$ axis.  The dimensionless parameters $A_x = e{\mathcal A}_x a/\hbar$ and $A_y = e{\mathcal A}_y a/\hbar$ are used to charaterize the driving field.

The separation vector may be expressed as $\left(\mathbf{R}_{n} - \mathbf{R}_m\right) = d_x \boldsymbol{\hat{x}} + d_y \boldsymbol{\hat{y}} + d_z \boldsymbol{\hat{z}}$. Substituting the dot product with $\mathbf{A}$ back into Eq.~\ref{eff-hop} and carrying out the integration yields  
\begin{equation}
\tilde{t}_{nm}^{p,q} = t_{nm} J_{p-q}(\Gamma) \mathrm{e}^{i(p-q)\Theta},
\label{eff-hop1}
\end{equation}
with  
\begin{eqnarray}
\Gamma &=& \sqrt{\left(A_xd_x\right)^2 + \left(A_yd_y\right)^2 + 2A_xA_yd_xd_y\cos{\phi}}\;\;\;\\
\Theta &=& \tan^{-1}{\left(\frac{A_yd_y\sin{\phi}}{A_xd_x + A_yd_y\cos{\phi}}\right)},
\end{eqnarray}     
where $J_{p-q}(\Gamma)$ is the Bessel function of the first kind of order $(p-q)$. Since $\Gamma$ depends on the orientation of the bond vector, the hopping amplitudes become bond-direction dependent in the presence of periodic driving.

\section{\label{appb}Band structure of extended SSH chain with zigzag pattern}
The knowledge of the electronic band structure in $k$-space is crucial for estimating the total bandwidth of the system under consideration. For an extended SSH chain arranged linearly, the computation of the $k$-space band structure becomes a straightforward task. Considering a unit cell with three sublattices $A$, $B$, and $C$, the $k$-space Hamiltonian takes the form of a $3\times 3$ matrix, whose eigenvalues are given by~\cite{adamatios}
\begin{equation}
E_\lambda(k) = 2\sqrt{-\frac{P}{3}}\cos{\left[\frac{1}{3}\arccos{\left( \frac{3Q}{2P}\sqrt{\frac{-3}{P}} \right)} - \frac{2\pi\lambda}{3}\right]},\label{lin-chain}
\end{equation}
where 
\begin{eqnarray}
P&=& -\left(t_1^2 + t_2^2 + t_3^2\right),\\
Q &=& -2t_1 t_2 t_3 \cos{k}.
\end{eqnarray}
Here $\lambda=0,1,2$. Thus, three distinct bands emerge for the linear extended SSH chain.

On the other hand, in our case to capture the zigzag geometry of an extended SSH chain, we consider a doubled (super) cell containing six-sites $\left[ A_1,B_1,C_1,A_2,B_2,C_2\right]$. All nearest-neighbor hoppings $t_1, t_2, t_3$ alternate their bond orientation between consecutive unit cells:
\begin{equation}
\bm{\delta}^{(+)} = \left(\frac{1}{2},+\frac{\sqrt{3}}{2}\right), \quad\quad\bm{\delta}^{(-)} = \left(\frac{1}{2},-\frac{\sqrt{3}}{2}\right),
\end{equation}
with the supercell lattice vector along $x$-direction $\mathbf{a}_{sc} = (a,0)$, and $a=1$.

The Fourier transformations are defined as
\begin{eqnarray}
c_{n,\alpha} &=& \frac{1}{\sqrt{3N}} \sum_k e^{i\mathbf{k}\cdot \mathbf{R}_n} c_{k,\alpha},\\
c_{n,\alpha}^\dagger &=& \frac{1}{\sqrt{3N}} \sum_k e^{-i\mathbf{k}\cdot \mathbf{R}_n} c_{k,\alpha}^\dagger,
\end{eqnarray}
where $k=\frac{2\pi m}{3N}$ with $m=0,1,\ldots,3N-1$ for periodic boundary conditions. Here, $n$ denotes the unit-cell index, $\alpha\in{A,B,C}$ labels the sublattice and $\mathbf{R}_n=2an\hat{x}$.

Substitute the Fourier transforms into the real-space Hamiltonian as given in Eq.~\ref{Eqn1} and following the standard procedure, the Hamiltonian in $k$-space in the basis $\left[ A_1,B_1,C_1,A_2,B_2,C_2\right]$ becomes
\begin{widetext}
\begin{equation}
H(k) =
\begin{pmatrix}
0 & t_1 e^{-i\mathbf{k}\cdot\bm{\delta}^{(+)}} & 0 & 0 & 0 & t_3 e^{-i\left(\mathbf{k}\cdot\bm{\delta}^{(-)}+\mathbf{a}_{sc}\right)}\\
t_1 e^{i\mathbf{k}\cdot\bm{\delta}^{(+)}} & 0 & t_2 e^{-i\mathbf{k}\cdot\bm{\delta}^{(-)}} & 0 & 0 & 0 \\
0 & t_2 e^{i\mathbf{k}\cdot\bm{\delta}^{(-)}} & 0 & t_3 e^{-i\mathbf{k}\cdot\bm{\delta}^{(+)}} & 0 & 0 \\
0 & 0 & t_3 e^{i\mathbf{k}\cdot\bm{\delta}^{(+)}} & 0 & t_1 e^{-i\mathbf{k}\cdot\bm{\delta}^{(-)}} & 0 \\
0 & 0 & 0 & t_1 e^{i\mathbf{k}\cdot\bm{\delta}^{(-)}} & 0 & t_2 e^{-i\mathbf{k}\cdot\bm{\delta}^{(+)}} \\
t_3 e^{-i\left(\mathbf{k}\cdot\bm{\delta}^{(-)}+\mathbf{a}_{sc}\right)} & 0 & 0 & 0 & t_2 e^{i\mathbf{k}\cdot\bm{\delta}^{(+)}} & 0
\end{pmatrix}.
\label{k-ham}
\end{equation}
\end{widetext}
Here the on-site potentials due to the applied bias are considered zero.

Diagonalizing over the reduced Brillouin zone $k_x \in \left[-\frac{\pi}{2a},\frac{\pi}{2a}\right]$ yeilds six bands $E_\lambda\left(k_x\right), \lambda=1,2,3,\ldots 6$, fully capturing the zigzag pattern. Another important point we shuld mention here is that although the zigzag geometry has a 2D appearance, the electrons hop strictly along the chain direction. Therefore, the band structure depends only on $k_x$, and we fix $k_y=0$ for all calculations.

Using the $k$-space Hamiltonian defined above, we present the band structure for both isotropic and anisotropic hopping cases, in the presence and absence of light, as shown in Fig.~\ref{band-structure}. The light irradiation is considered in the high-frequency regime, as described in the Results section, where the hopping amplitudes are renormalized by the zeroth-order Bessel function.

In the absence of light, the isotropic case with $t_1 = t_2 = t_3 = 1\,$eV exhibits six distinct energy levels, as shown in Fig.~\ref{band-structure}(a). 
\begin{figure}[ht!]
\includegraphics[width=0.238\textwidth]{Fig8a.eps}\hfill\includegraphics[width=0.238\textwidth]{Fig8b.eps}\vskip 0.1 in
\includegraphics[width=0.238\textwidth]{Fig8c.eps}\hfill\includegraphics[width=0.238\textwidth]{Fig8d.eps}\\
\caption{(Color online.) Band structure of the extended SSH chain with a zigzag pattern. (a,b) Results in the absence of light. (c,d) Results in the presence of light. (a,c) Isotropic case with $t_1=t_2=t_3=1,$eV. (b,d) Anisotropic case with $t_1=1\,$eV, $t_2=1.25\,$eV, and $t_3=1.5\,$eV. The light parameters are $A_x=A_y=1$ and $\phi=0$.}
\label{band-structure}
\end{figure}
For the anisotropic case, with $t_1 = 1\,$eV, $t_2 = 1.25\,$eV, and $t_3 = 1.5\,$eV, the band structure forms three separate bands, reflecting the extended SSH chain with a zigzag pattern and unequal hopping amplitudes, as seen in Fig.~\ref{band-structure}(b). This three-band splitting is a characteristic signature of the extended SSH chain~\cite{adamatios}.

Under irradiation, the isotropic case shows a pronounced modification of the band structure, exhibiting two apparent bands, as shown in Fig.~\ref{band-structure}(c). This splitting arises from the zigzag geometry: the two directional bonds lead to two distinct renormalized hopping amplitudes when the bare hoppings are isotropic. In the anisotropic case, irradiation also alters the band structure compared to the absence-of-light scenario, as illustrated in Fig.~\ref{band-structure}(d).

Since the directional bonds contribute to the phase factor in the hopping, the band width in the absence of light can be estimated from Eq.~\ref{lin-chain} as
\begin{equation}
-2\sqrt{\frac{t_1^2 + t_2^2 +t_3^2}{3}} \quad\text{to} \quad 2\sqrt{\frac{t_1^2 + t_2^2 +t_3^2}{3}}.
\end{equation}

\section{\label{appc}Justification of high frequency regime}
The Floquet Hamiltonian in Sambe space acts on the basis
$\{\,\lvert u_{p,\alpha}(\mathbf{k})\rangle\,\}$
where $p$ indexes Floquet replicas (Fourier harmonics) and $\alpha$ labels the sublattice space. Restricting to three replicas $p=\{-1,0,+1\}$ (ordered as $-1,0,+1$) the Floquet Hamiltonian acquires a block structure~\cite{light-new1,light-new2,light-new3,light-new4}
\begin{equation}
\mathcal{H}_{\mathbf{k}}^{(F)} =
\begin{pmatrix}
H_0(\mathbf{k}) - \hbar\Omega\,I & H_{-1}(\mathbf{k}) & 0 \\
H_{+1}(\mathbf{k}) & H_0(\mathbf{k}) & H_{-1}(\mathbf{k}) \\
0 & H_{+1}(\mathbf{k}) & H_0(\mathbf{k}) + \hbar\Omega\,I
\end{pmatrix},
\label{floquet-3rep}
\end{equation}
where each entry is a matrix acting on the sublattice space and $I$ is the identity ($6\times 6$) in that space.
The general block element connecting replica $p$ to $q$ is
\begin{equation}
\big[ \mathcal{H}_{\mathbf{k}}^{(F)}\big]_{p,q} \equiv H_{p-q}(\mathbf{k}) + p\hbar\Omega\,\delta_{p,q}\,I,
\end{equation}
with $H_m(\mathbf{k})$ the $m$th Fourier component of the time-periodic Hamiltonian:
\begin{equation}
H_m(\mathbf{k}) \;=\; \frac{1}{\mathbb{T}}\int_0^\mathbb{T} H(\mathbf{k},\tau)\, e^{-im\Omega\tau}\,d\tau.
\end{equation}
Using the minimal coupling result from Eq.~(\ref{eff-hop1}), each hopping contribution to $H_m(\mathbf{k})$ acquires a Bessel renormalization:
\begin{equation}
H_m(\mathbf{k}) \;=\; \sum_{\boldsymbol{d}} t_{\boldsymbol{d}}\; J_{m}(\Gamma_{\boldsymbol{d}})\, e^{i m \Theta_{\boldsymbol{d}}}\, e^{i\mathbf{k}\cdot\boldsymbol{d}}\, S_{\boldsymbol{d}},
\label{Hm-k}
\end{equation}
where $\boldsymbol{d}=\mathbf{R}_n-\mathbf{R}_m$ runs over bond vectors, $S_{\boldsymbol{d}}$ is the matrix in the sublattice basis that implements the hop along $\boldsymbol{d}$, and $\Gamma_{\boldsymbol{d}},\Theta_{\boldsymbol{d}}$ are defined in Eqs.~\ref{gam} and \ref{thet}.

The explicit form of $H_m(\mathbf{k}) (m=0,\pm 1)$ for the extended SSH chain with zigzag pattern can be written from Eq.~\ref{k-ham} as
\begin{widetext}
\begin{equation}
H_m(\mathbf{k}) =
\begin{pmatrix}
0 & F_m^{1,1} & 0 & 0 & 0 & F_m^{2,3} \\
 \left( F_m^{1,1}\right)^*  & 0 &  F_m^{1,2}  & 0 & 0 & 0 \\
0 & \left( F_m^{1,2}\right)^*  & 0 & F_m^{1,3} & 0 & 0 \\
0 & 0 & \left( F_m^{1,3}\right)^* & 0 & F_m^{2,1} & 0 \\
0 & 0 & 0 & \left( F_m^{2,1}\right)^*  & 0 & F_m^{2,2} \\
\left( F_m^{2,3}\right)^*  & 0 & 0 & 0 & \left( F_m^{2,2}\right)^*  & 0
\end{pmatrix},
\label{k-replica}
\end{equation}
\end{widetext}
where
\begin{eqnarray}
F_m^{1,1} &=& t_1 \,J_{m}(\Gamma_{\bm{\delta}^{(+)}})\, e^{i m \Theta_{\bm{\delta}^{(+)}}}\, e^{-i\mathbf{k}\cdot\bm{\delta}^{(+)}},\\
F_m^{1,2} &=& t_2  \,J_{m}(\Gamma_{\bm{\delta}^{(-)}})\, e^{i m \Theta_{\bm{\delta}^{(-)}}}\, e^{-i\mathbf{k}\cdot\bm{\delta}^{(-)}},\\
F_m^{1,3} &=& t_3  \,J_{m}(\Gamma_{\bm{\delta}^{(+)}})\, e^{i m \Theta_{\bm{\delta}^{(+)}}}\, e^{-i\mathbf{k}\cdot\bm{\delta}^{(+)}},\\
F_m^{2,1} &=& t_1 \,J_{m}(\Gamma_{\bm{\delta}^{(-)}})\, e^{i m \Theta_{\bm{\delta}^{(-)}}}\,  e^{-i\mathbf{k}\cdot\bm{\delta}^{(-)}},\\
F_m^{2,2} &=& t_2 \,J_{m}(\Gamma_{\bm{\delta}^{(+)}})\, e^{i m \Theta_{\bm{\delta}^{(+)}}}\,  e^{-i\mathbf{k}\cdot\bm{\delta}^{(+)}},\\
F_m^{2,3} &=& t_3 \,J_{m}(\Gamma_{\bm{\delta}^{(-)}})\, e^{i m \Theta_{\bm{\delta}^{(-)}}}\, e^{-i\left(\mathbf{k}\cdot\bm{\delta}^{(-)}+\mathbf{a}_{sc}\right)},
\end{eqnarray}
and $\left(F_m^{i,j}\right)^*$ is the complex conjugate of $F_m^{i,j}$. The indices $i$ and $j$ denote the unit-cell index of the supercell, which consists of two unit cells, and the sublattice index, respectively.

\begin{figure}[ht!]
\includegraphics[width=0.238\textwidth]{Fig9a.eps}\hfill\includegraphics[width=0.238\textwidth]{Fig9b.eps}\vskip 0.1 in
\includegraphics[width=0.238\textwidth]{Fig9c.eps}\hfill\includegraphics[width=0.238\textwidth]{Fig9d.eps}\\
\caption{(Color online.) Quasi-energy spectrum as a function of $k_x$ for the extended SSH chain with three Floquet replicas ($m=0,\pm 1$). The hopping amplitudes are isotropic with $t_1=t_2=t_3=1\,$eV in panels (a,c), and anisotropic with $t_1=1\,$eV, $t_2=1.25\,$eV, and $t_3=1.5\,$eV in panels (b,d). The upper row corresponds to $\hbar\Omega=0.5\,$eV, and the lower row to $\hbar\Omega=6\,$eV. The light parameters are $A_x=A_y=1$ and $\phi=\pi/3$.}
\label{fig-k-space-replica}
\end{figure}

We plot the band structure of the system considering three Floquet replicas ($m=0,\pm 1$), as shown in Fig.~\ref{fig-k-space-replica} by diagonalizing the matrix as given in Eq.~\ref{k-replica}. The upper row corresponds to low-frequency driving with $\hbar\Omega=0.5\,$eV, while the lower row corresponds to high-frequency driving with $\hbar\Omega=6\,$eV. In other words, the upper panels represent the case where the photon energy is smaller than the bandwidth of the undriven system, whereas the lower panels correspond to photon energy larger than the bandwidth. Both isotropic and anisotropic hopping configurations are considered.

In the low-frequency regime (Figs.~\ref{fig-k-space-replica}(a,b)), the Floquet replicas ($m=0,\pm1$) strongly overlap and hybridize in both isotropic and anisotropic cases, with the anisotropic case exhibiting a slightly broader energy window. By contrast, in the high-frequency regime (Figs.~\ref{fig-k-space-replica}(c,d)), the Floquet replicas are well separated in energy. The $m=0$ replica (the central band) corresponds to the renormalized or ``dressed" version of the original static band structure, while the $m=\pm 1$ replicas appear as simple sidebands shifted by $\pm\hbar\Omega$ without interacting with the central band. This clear separation justifies neglecting higher-order Floquet replicas in the high-frequency regime.

\textbf{Real-space representation.} Similar to the $k$-space, in real space (site index $r$ and replica index $p$), the Floquet Hamiltonian with three replicas takes the form
\begin{equation}
\mathcal{H}^{(F)} =
\bordermatrix{ & (r,-1) & (r,0) & (r,+1) \cr
(r^\prime,-1) & H_0^{rr^\prime} - \hbar\Omega\,I & H_{-1}^{rr^\prime} & 0 \cr
(r^\prime,0)  & H_{+1}^{rr^\prime} & H_0^{rr^\prime} & H_{-1}^{rr^\prime} \cr
(r^\prime,+1) & 0 & H_{+1}^{rr^\prime} & H_0^{rr^\prime} + \hbar\Omega\,I
},
\end{equation}
where
\begin{equation}
H_m^{rr'} \;=\; \frac{1}{\mathbb{T}}\int_0^\mathbb{T} t_{rr'}(\tau)\, e^{-im\Omega\tau}\, d\tau
\;=\; t_{rr'}\, J_m(\Gamma_{rr'})\, e^{i m\Theta_{rr'}}.
\end{equation}
Thus in real space each off-diagonal Floquet block $H_m^{rr'}$ connects the electronic amplitude on site $r'$ in replica $q$ to site $r$ in replica $p$ with $m=p-q$. Here $I$ an identity ($3N\times 3N$), where $N$ is the number of unit cells in real space. 

Considering two unit cells, the explicit form of $H_m^{rr'} (m=0,\pm 1)$ takes the form
\begin{equation}
H_m =
\begin{pmatrix}
0 & G_m^{1,2} & 0 & 0 & 0 & 0 \\
 G_m^{2,1}  & 0 &  G_m^{2,3}  & 0 & 0 & 0 \\
0 & G_m^{3,2}  & 0 & G_m^{3,4} & 0 & 0 \\
0 & 0 & G_m^{4,3} & 0 & G_m^{4,5} & 0 \\
0 & 0 & 0 & G_m^{5,4}  & 0 & G_m^{5,6} \\
0  & 0 & 0 & 0 & G_m^{6,5}  & 0
\end{pmatrix},
\label{real-replica}
\end{equation}
where
\begin{eqnarray}
G_m^{1,2} &=& t_1 \,J_{m}(\Gamma_{\bm{\delta}^{(+)}})\, e^{i m \Theta_{\bm{\delta}^{(+)}}},\\
G_m^{2,3} &=& t_2  \,J_{m}(\Gamma_{\bm{\delta}^{(-)}})\, e^{i m \Theta_{\bm{\delta}^{(-)}}},\\
G_m^{3,4} &=& t_3  \,J_{m}(\Gamma_{\bm{\delta}^{(+)}})\, e^{i m \Theta_{\bm{\delta}^{(+)}}},\\
G_m^{4,5} &=& t_1 \,J_{m}(\Gamma_{\bm{\delta}^{(-)}})\, e^{i m \Theta_{\bm{\delta}^{(-)}}},\\
G_m^{5,6} &=& t_2 \,J_{m}(\Gamma_{\bm{\delta}^{(+)}})\, e^{i m \Theta_{\bm{\delta}^{(+)}}},\\
G_m^{r^\prime,r} &=& \left(G_m^{r,r^\prime}\right)^*.
\end{eqnarray}
The indices $r$ and $r^\prime$ denote the site indices.

For completeness, we plot the energy eigenvalues of the real-space Floquet Hamiltonian as a function of the driving energy $\hbar\Omega$ as shown in Fig.~\ref{fig-real-space-replica}. We consider the extended SSH chain consisting of 7 unit cells, i.e., a total of 21 sites as in the Results section. Accordingly, each Floquet block $H_m^{rr^\prime}$ is a $21\times 21$ matrix. As in the $k$-space case, we restrict the calculation to three Floquet replicas, $m=0,\pm 1$. Since the real-space Floquet Hamiltonian can naturally incorporate on-site potentials due to an applied bias, we also include this effect.

\begin{figure}[ht!]
\includegraphics[width=0.238\textwidth]{Fig10a.eps}\hfill\includegraphics[width=0.238\textwidth]{Fig10b.eps}\vskip 0.1 in
\includegraphics[width=0.238\textwidth]{Fig10c.eps}\hfill\includegraphics[width=0.238\textwidth]{Fig10d.eps}\\
\caption{(Color online.) Energy eigenvalues of the real-space Floquet Hamiltonian as a function of driving energy $\hbar\Omega$ for an extended SSH chain with 7 unit cells (21 sites). Three Floquet replicas ($m=0,\pm1$) are included. Upper panels: $\mathcal{V}=0$ and lower panels: $\mathcal{V}=1\,$V. (a,c) $A_x=A_y=1$, $\phi=0$ and (b,d) $A_x=A_y=1$, $\phi=\pi/3$. }
\label{fig-real-space-replica}
\end{figure}

The upper panels correspond to zero applied bias ($\mathcal{V}=0$), while the lower panels correspond to a finite bias of $\mathcal{V}=1\,$V. Two sets of light parameters are considered: $A_x=A_y=1$, $\phi=0$ (Figs.~\ref{fig-real-space-replica}(a,c)), and $A_x=A_y=1$, $\phi=\pi/3$ (Figs.~\ref{fig-real-space-replica}(b,d)). For $\phi=0$, the overlap between Floquet replicas persists up to $\hbar\Omega\approx 4.5\,$eV (Fig.~\ref{fig-real-space-replica}(a)), while for $\phi=\pi/3$ the overlap region extends only up to $\hbar\Omega\approx 2\,$eV (Fig.~\ref{fig-real-space-replica}(b)) in the absence of bias. The role of the applied bias is particularly evident in the low-energy regime, where without bias, the eigenvalues are nearly constant, forming a rectangular strip, while the inclusion of bias breaks this structure, as seen in the lower panels. Moreover, the overlap region is slightly extended in Figs.~\ref{fig-real-space-replica}(c,d) compared to their unbiased counterparts.

Overall, the results confirm that with increasing driving energy the Floquet replicas progressively separate. Hence, in the high-frequency regime it is sufficient to retain only the zeroth-order Floquet band. For our analysis, $\hbar\Omega=6\,$eV is an appropriate choice for the high-frequency regime.


\end{document}